\documentclass[aps,prl,twocolumn,groupedaddress,10pt]{revtex4}

\usepackage[english]{babel}
\usepackage[T1]{fontenc}
\usepackage{graphicx}
\usepackage{color} 
\usepackage{bbm}
\usepackage{amsmath}
\usepackage{hyperref}
\hypersetup{pdfstartview={FitH}}

\newcommand{\bq}{\begin{equation}}
\newcommand{\eq}{\end{equation}}
\newcommand{\ba}{\begin{eqnarray}}
\newcommand{\ea}{\end{eqnarray}}

\begin{document} 

\title{Evolutionary adaptation is facilitated by the presence of lethal genotypes}
\author{Viktoryia Blavatska$^{1,2}$ and Bartlomiej Waclaw$^{1,3}$}
\affiliation{
$^1$ Dioscuri Centre for Physics and Chemistry of Bacteria, Institute of Physical Chemistry PAS, Kasprzaka 44/52, 01-224 Warsaw, Poland\\
$^2$ Institute for Condensed Matter Physics of the National Academy of Sciences of Ukraine,\\
79011 Lviv, Ukraine\\
$^3$School of Physics and Astronomy, The University of Edinburgh, JCMB, Peter Guthrie Tait Road, Edinburgh
EH9 3FD, United Kingdom \\
}

% below 600 chars!
\begin{abstract}
The adaptation rate in theoretical models of biological evolution increases with the mutation rate but only to a point when mutations into lethal states cause extinction. One would expect that removing such states should be beneficial for evolution. We show here that, counter-intuitively, lethal mutations speed up adaptation on rugged fitness landscapes with many fitness maxima and minima, if strong competition for resources exist. We consider a modified stochastic version of the quasispecies model with two types of genotypes, viable and lethal, and show that increasing the rate of lethal mutations decreases the time to evolve the best-fit genotype. This can be explained by an increased rate of crossing fitness valleys, facilitated by reduced selection against less-fit variants. 
\end{abstract}

\maketitle

Biological evolution is often thought to be a slow process occurring over timescales of 10k-100ks years \cite{endicott_evaluating_2009}. However, ecological speciation which causes a species to split into non-interbreeding sub-populations can occur in just tens of generations in animals and plants
\cite{hendry_speed_2007}, and microevolution - a significant change in allele (gene variants) frequencies can occur over a few hours in microbes exposed to antibiotics \cite{zhang_acceleration_2011,toprak_evolutionary_2011,greulich_mutational_2012,hermsen_rapidity_2012}.

The rate of biological evolution is affected, among others, by the mutation rate \cite{nowak_evolutionary_2006}, %\cite{ref1,13_21_from_Clune2008},
the structure of the fitness landscape \cite{Weinreich2005,Weissman2009,Park2009,tan_hidden_2011,szendro_quantitative_2013}, its temporal dynamics \cite{Kashtan2007,Mustonen2008,ashcroft_fixation_2014,taitelbaum_population_2020},  
and spatial structure of the population \cite{Martens2011,Hermsen2012,Greulich2012}.
Here we focus on its dependence on the fitness landscape (FL) and the mutation rate, which has been studied in the quasispecies model \cite{eigen_principle_1977}.
It has been shown that the time it takes biological evolution to reach the best-adapted genotype decreases with increasing mutation rate \cite{Campos2002,jain_deterministic_2006,jain_adaptation_2007,jain_evolutionary_2007}. From the physics viewpoint, this is expected because the mutation rate plays a similar role to the temperature in a quantum-spin system with an energy landscape given by the inverted fitness landscape \cite{baake_mutation-selection_2001}. However, if the mutation rate is too high, the population becomes delocalized (``error catastrophe'') and may even die out if the fitness landscape has a sufficient density of genotypes with null or negative fitness \cite{tejero_effect_2010,woo_quantitative_2012}. In real fitness landscapes, such genotypes make up a substantial fraction of all genotypes \cite{eyre_walker_distribution_2007}. It is therefore not possible to make the mutation rate arbitrarily large and, consequently, there is a limit on how fast evolution can be on a given fitness landscape.

In some situations such as directed evolution of proteins \cite{Arnold1998,Packer2015}, one would however wish to increase the rate of evolution. Since the presence of lethal genotypes is a limiting factor, one could think that reducing the fraction of such genotypes, e.g., by changing experimental conditions, should be beneficial. %However, increasing the mutation rate poses a problem that eventually too many lethal genotypes will be generated and the population will go extinct.

\begin{figure}[b!]
	\begin{center}
	  \includegraphics[width=\columnwidth]{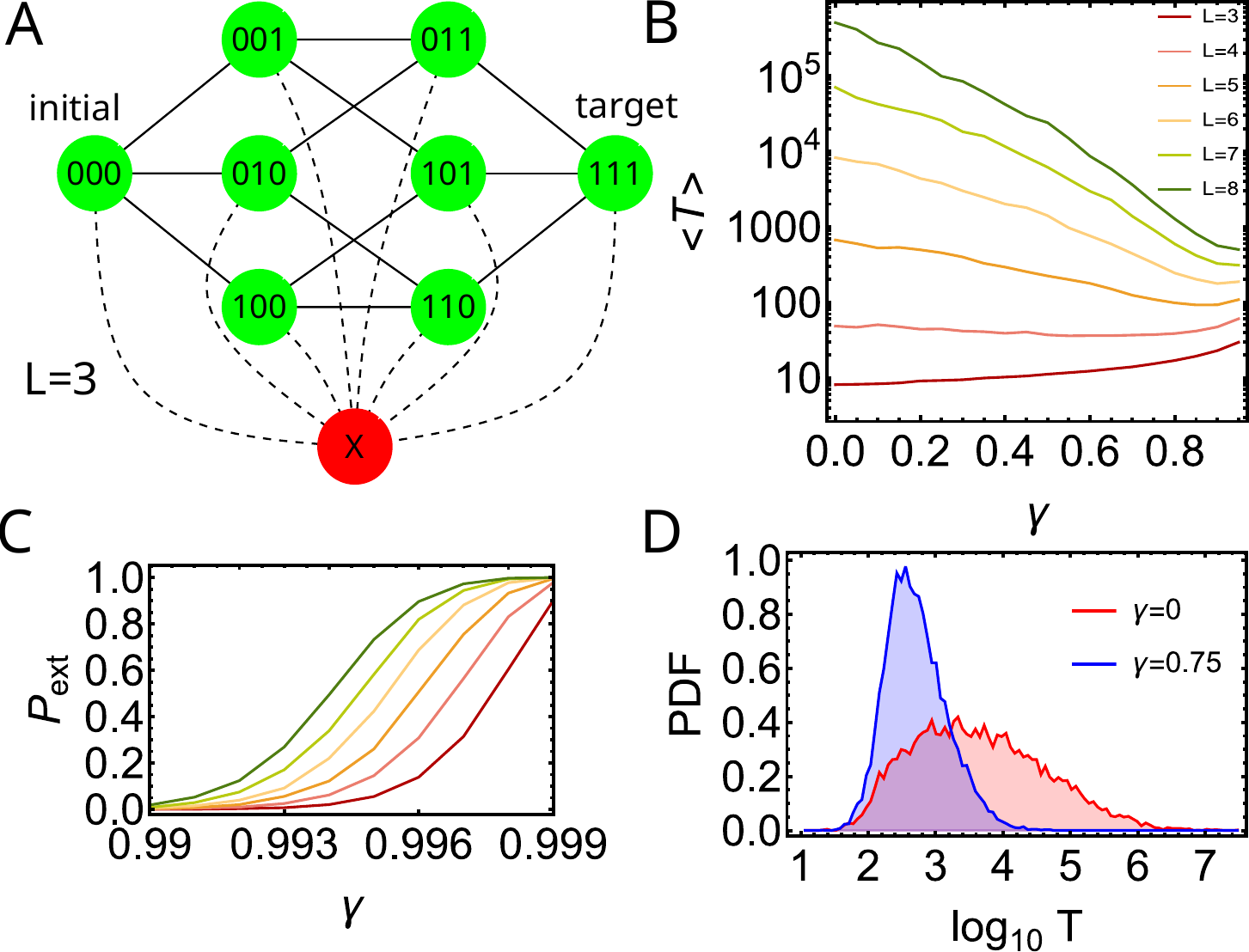}
		\caption{(A) The genotype space of the model for $L=3$. Green = viable states, red = lethal state. Lines represent mutations. (B) Mean adaptation time $\langle T \rangle$ decreases with increasing  probability $\gamma$ of lethal mutations, for $L\geq 5$. Colors = different $L$. (C) The extinction probability $P_{\rm ext}$ is low even for $\gamma\approx 0.99$. (D) The distribution of adaptation time $P(T)$, for $\gamma=0$ and $\gamma=0.75$. $L=7$ in both cases. %\red{BW: add aver. pop. size plot?}
  }
\label{fig:1}	
\end{center}
\end{figure}

%In the quasispecies model, increased rate of evolution is caused by the increased transition probability between the genotypes of the fitness landscape.

To explore how the presence of non-viable (lethal) genotypes affects the rate of evolution, we consider a stochastic version of the quasispecies model on a fitness landscape with viable and non-viable genotypes. In contrast to previous works on the quasispecies model, we divide up the genotype space into two regions: one for genes relevant for adaptation to a new environment, and another one for housekeeping genes crucial for metabolism, DNA replication, etc., which have already been highly optimized and therefore any mutation in them is likely to be lethal. This scenario is biologically more realistic than assigning zero fitness to random genotypes in the landscape.

%The quasispecies and related models focus only on a small set of genes. Here we wish to consider a more realistic scenario, in which mutations may occur also in other parts of the genome which we don't model explicitly. 

%Say we look at $L$ genomic loci, but there is $M\gg L$ loci which can also mutate and these mutations are lethal.
%If the per-loci mutation probability is $u$, then upon replication, an organism will much more likely mutate into the lethal state than the viable state.
%We now increase the proportion of lethal genotypes while keeping the mutation rate the same. How would the adaptation time be affected?

% -----------model------------------
To be specific, we consider a population of organisms replicating and dying stochastically. Each organism has a genotype $i=0,\dots,2^L-1$ represented by a binary sequence of length $L$.
The state of the system is described by a vector $\{n_i\}$, where $n_i$ is the number of organisms of type $i$.
An organism of type $i$ replicates with rate $f_i$ and dies with rate $N/K$, where $N=\sum_i n_i$, and $K$ is the (soft) carrying capacity of the system. Upon replication, the organism either produces a copy of itself or a mutant. The mutant can be either viable, or non-viable (lethal). We assume that faithful replication occurs with probability $1-\mu-\gamma$, a viable mutant is generated with probability $\mu$, and a lethal mutant with probability $\gamma$. A lethal mutant is instantaneously removed from the system. The genotype of a viable mutant is obtained by inverting a randomly selected letter (0 or 1) of the binary representation of $i$. The genotype space is therefore an $L$-dimensional hypercube, with an additional node representing the lethal state connected to all other genotypes (see Fig. \ref{fig:1}A).

%\begin{figure}[b!]
	%\begin{center}
	  %\includegraphics[width=70mm]{times.eps}
 %\includegraphics[width=70mm]{timesflat.eps}	
 % \caption }
%\label{times}	
%\end{center}
%\end{figure}

%We may think of $\mu$ and $\gamma$ as representing mutations in different parts of organism's genome: $\mu$ is the rate of mutations that affect genes relevant for adaptation to a new environment, whereas $\gamma$ affects "housekeeping" genes crucial for metabolism, DNA replication, etc., which have already been highly optimized and therefore any mutation in them is likely to be lethal.

The replication rates $\{f_i\}$ are drawn independently from the uniform distribution on $[0,1)$, except for $f_0=0.5$ and $f_{2^L-1}=1$ which we fix so that $i=2^L-1$ is always the fittest genotype, and the initial genotype has intermediate fitness. %Evolution will therefore proceed towards the global fitness maximum at $i=2^L-1$.
The system is initialized with $K$ organisms of genotype $i=0$ and simulated until at least one organism of the fittest genotype $i=2^L-1$ emerges. 

For $\gamma=0$, the model is essentially the stochastic quasispecies model with a maximally-rugged fitness landscape \cite{park_quasispecies_2010}. %It is known that in this model, evolution is the fastest, the larger $\mu$ becomes.
%say why mu cannot be too large
For $\gamma>0$, the fitness landscape contains a fraction $\gamma/(\mu+\gamma)$ of lethal genotypes. By construction, $\mu+\gamma$ must be smaller than one since it is the total mutation probability. We cannot therefore make either $\mu$ or $\gamma$ too large.

We are interested in the large-population/large-mutation rate regime because we want evolution to be fast. In this limit, low-fitness variants do not fix in the population, and therefore the rate of evolution decreases monotonously with population size \cite{jain_evolutionary_2011}. Since the presence of lethal genotypes reduces the population size,
%A naive expectation would be then that  and hence the number of events that can lead to successful (non-lethal) mutations will also be reduced.
evolution is expected to be slower for $\gamma>0$.
We shall see that this expectation is not always correct, and that lethal genotypes, while not participating in the evolutionary process, have a significant effect on the rate of evolution.

%-----------results-----------------
To compare the model with and without lethal genotypes, we simulated 1000 copies of the model for $K=1000,\mu=0.04$, $L=3,\dots,8$ and a range of $\gamma=0,\dots,0.96$, and measured the time $T$ it took for a single organism of best-adapted genotype to evolve. 
We used the same sequence of randomly generated FLs for each $\gamma$. Figure \ref{fig:1}B shows that the average adaptation time increases exponentially with $L$. %\red{BW: increase too slow for L=9} 
However, the increase is slower in the model with $\gamma>0$. In fact, for $L>4$, the presence of the lethal genotype speeds up evolution. %\red{BW: Explain later why $L=4$ is critical} 
For $L=8$ and $\gamma=0.75$, evolution is three orders of magnitude faster than for $\gamma=0$. %In the SI we show the same effect for the Moran-like model with a stiff cap on the total number of organisms. \red{DO?} % write also about the model in which replication and mutations are decoupled?

This behaviour is counter-intuitive since the total number of organisms in the system is significantly reduced for $\gamma>0$; for example, for $\gamma=0.75$ the population reaches only about $20\%$ of the maximum capacity. Although the per-capita birth rate does not depend on population size, the average rate with which mutants are generated is lower because there are fewer births. Moreover, the effect shows up only on large-enough fitness landscapes; for $L<4$, non-lethal model is actually faster. The lethal-genotype model is slower for any $L$ also on a flat landscape (all fitnesses the same, results not shown), on which there is only genetic drift and no selection (except for non-lethality). The effect is also unrelated to the increased probability of extinction in the lethal model, since extinction becomes likely only for $\gamma$ very close to $1$  (Fig. \ref{fig:1}C); in this limit Fig. \ref{fig:1}B shows that $\langle T\rangle$ increases slightly with $\gamma$.

In what follows we shall focus on two cases: $\gamma=0$ (``non-lethal'') and $\gamma=0.75$ (``lethal''), for which the extinction probability is effectively zero for $K=1000$. The choice of $\mu=0.04,\gamma=0.75$ may be interpreted as about 1 in 20 mutations being viable, the rest being lethal. %It also gives the total probability of $0.2$ of faithful replication, i.e., almost all replication events produce mutants. This is close to the extinction threshold and we expect the evolution to be fastest in this regime.
Figure (\ref{fig:1}D) shows the distribution of adaptation times for $L=7$. While the distributions of the adaptation time are very broad for both models (as expected due to the stochastic nature of the evolutionary process), we notice that the non-lethal model has a distribution skewed to the right, which contributes to the much longer average adaptation time.

To understand whether the two models differ only for certain fitness landscapes or all landscapes, we set $\mu=0.04$, and $\gamma=0$ (no lethal genotypes) or $\gamma=0.75$ (lethal genotypes present), and run the models 20 times on 1000 random landscapes (identical for both models). We then looked at evolutionary pathways in both models for 100 landscapes with largest/smallest differences in the adaptation time. We shall call such landscapes ``slow'' and ``fast'', respectively.

\begin{figure}[t!]
	\begin{center}
	  \includegraphics[width=\columnwidth]{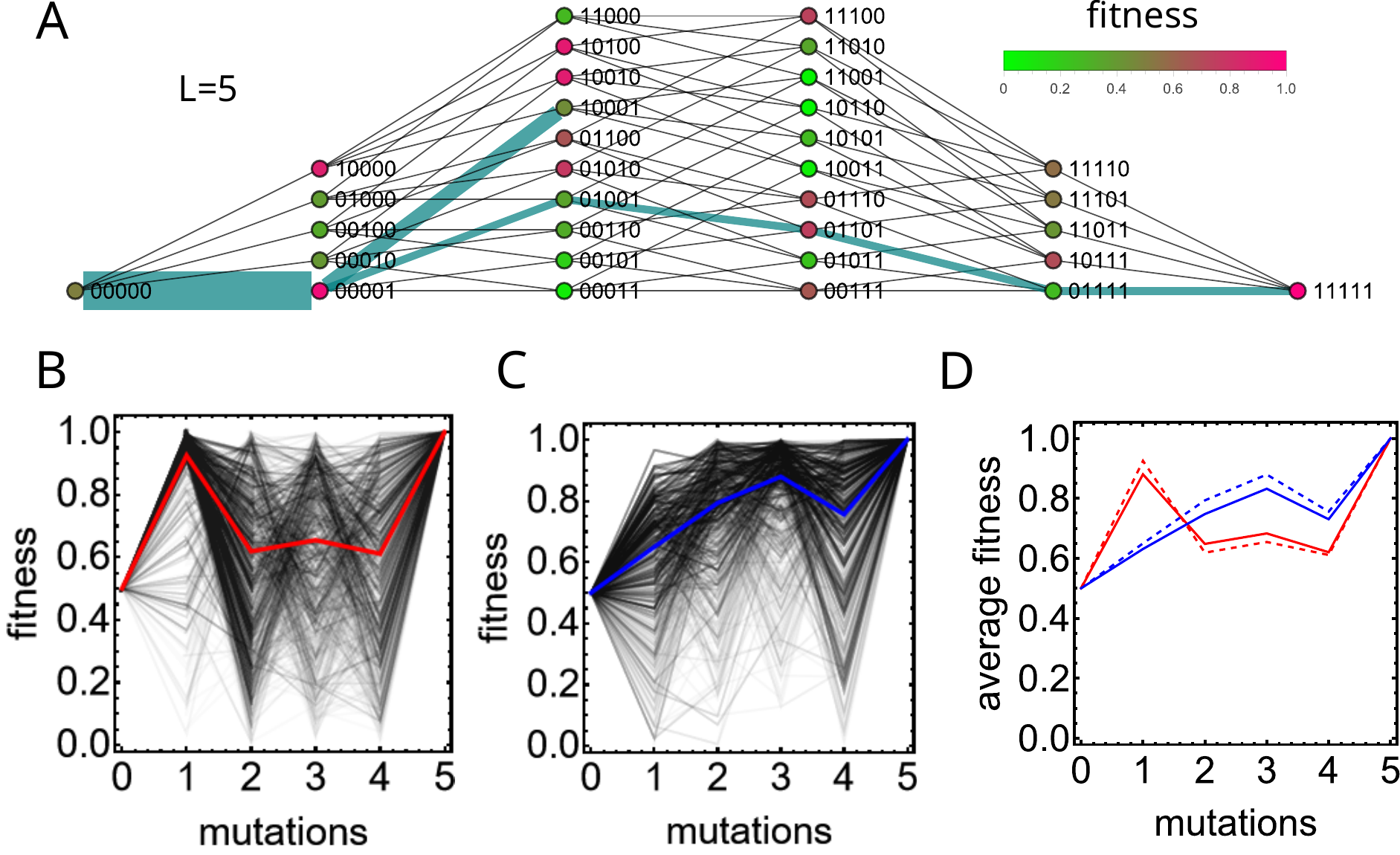}
	\caption{\label{fig:2}Evolutionary trajectories for $L=5$. (A) Example pathway in the non-lethal model for $\mu=0.04,\gamma=0$. (B-C) Fitness along evolutionary pathways, for the slowest 10\% of 1000 random FLs (panel B) and 10\% fastest FLs (panel C). Red and blue thick lines are averages over individual trajectories. (D) Average fitness along adaptive trajectories for slow (red) and fast (blue) FLs, for $\gamma=0$ (solid) and $\gamma=0.75$ (dashed). }
\end{center}
\end{figure}

Figure \ref{fig:2}A shows an example pathway in the non-lethal model on a ``slow'' landscape. The first mutation increased fitness to approx. the same value as the fitness of the final genotype, but subsequent mutations led through a fitness valley. This is typical for ``slow'' landscapes (Fig. \ref{fig:2}B). In contrast, evolutionary pathways on ``fast'' landscapes usually involve a more gradual fitness increase, with fewer and shallower fitness valleys (Fig. \ref{fig:2}C). Interestingly, the average fitness profile along the evolutionary trajectory looks very similar for the lethal and non-lethal model (Fig. \ref{fig:2}D). This suggests that the evolutionary speed-up provided by lethal genotypes does not rely on following different pathways, but rather on the enhanced rate of crossing of fitness valleys.

%how fitness changes along the pathways. that there is little difference between the two models, but that their evolutionary pathways involve multi-genotype wide fitness valleys on "slow" landscapes.

\begin{figure}[t!]
	\begin{center}
	  \includegraphics[width=\columnwidth]{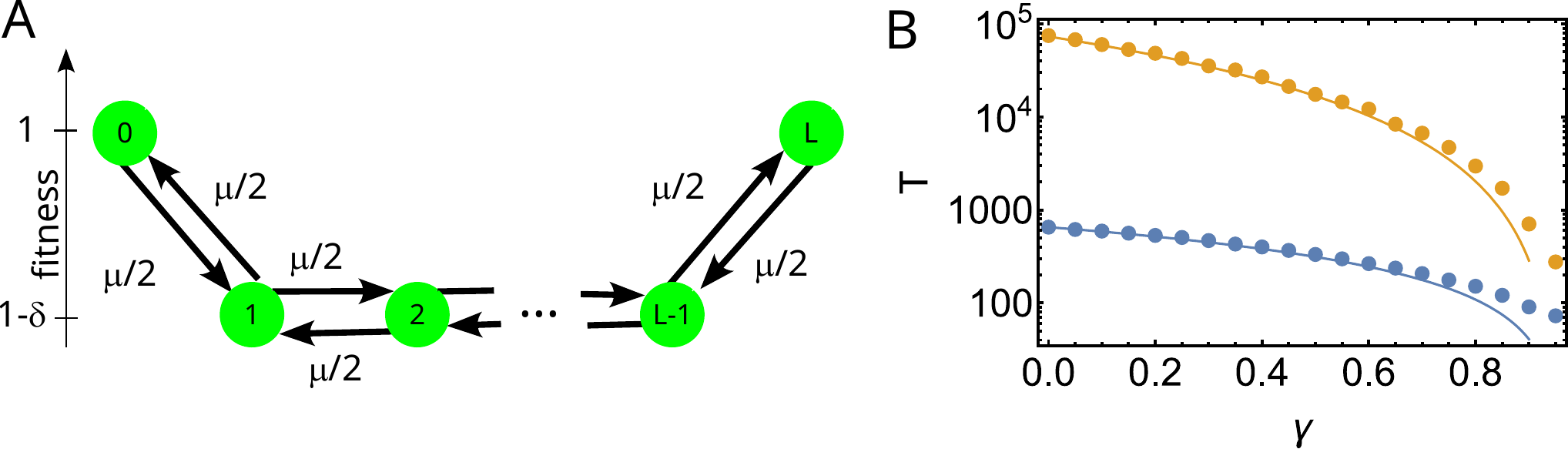}
		\caption{\label{fig:3}Evolution in a one-dimensional model with a wide fitness valley. (A) Schematic representation of the fitness landscape. (B) Time to reach genotype $i=L$. Points = simulation, line = theoretical predictions from Eq. (\ref{eq:1dmodel}): $L=3$ (blue), $L=4$ (yellow).  }
\end{center}
\end{figure}

To test this hypothesis, we investigated a simpler model with a one-dimensional fitness landscape (Fig. \ref{fig:3}A). Mutations can only change genotype $i$ to $i\pm 1$ with probability $\mu/2$ in either direction, and the fitness values are $f_0=1, f_1=f_2=...=f_{L-1}=1-\delta,f_L=1$, i.e., a flat fitness valley of depth $\delta$ separates the initial and the final genotype. This is meant to represent a single evolutionary pathway from Fig. \ref{fig:2} and is a special case of a more generic problem considered in Ref. \cite{weissman_rate_2009}, which significantly simplifies mathematical analysis. Numerical simulations confirmed that indeed the lethal-genotype model leads to much faster adaptation time on this fitness landscape (Fig. \ref{fig:3}B).

To understand this, we considered a deterministic counterpart of the stochastic model. Neglecting stochastic fluctuations, the abundances $\{n_i\}$ of organisms evolve according to the following set of equations: 
\ba
\frac{dn_0}{dt} &=& n_0 f_0 (1-\gamma-\mu) + \mu n_{1} f_{1}  - n_0 N(t)/K,  \label{eq:1} \\
% \frac{dn_1}{dt} &=& n_1 f_1 (1-\gamma-\mu) + \mu n_{0} f_{0} + (\mu/2) n_2 f_2 \nonumber\\ 
%&-& n_1 N(t)/K,  \label{eq:2} \\
 \frac{dn_i}{dt} &=& n_i f_i (1-\gamma-\mu) + (\mu/2)(n_{i+1} f_{i+1} + n_{i-1} f_{i-1}) \nonumber\\
&-& n_i N(t)/K,  
\ea
where $N(t)=\sum_i n_i(t)$. 
Assume for a while that the final state has fitness zero, so it acts as an absorbing boundary. The system of equations admits then a steady state solution, with abundances determined by the non-linear set of equations,
\ba
	0 &=& n_0 f_0 (1-\gamma-\mu) + \mu n_{1} f_{1} - n_0 N/K , \\
	0 &=& n_i f_i (1-\gamma-\mu) + (\mu/2)(n_{i+1} f_{i+1} + n_{i-1} f_{i-1}) \nonumber\\
&-& n_i N/K  ,
\ea
with $N=\sum_i n_i$. These equations will also correctly describe the quasi-stationary distribution for the case of non-zero fitness at $i=L$, as long as the transition rate from $i=L-1$ to $i=L$ (proportional to $\mu$) is small. For small $\mu$, we expect $n_{i+1}\ll n_i$, which enables us to write
\bq
	n_i \approx n_{i-1} \frac{f_{i-1}\mu/2}{N/K -f_i (1-\gamma-\mu)} .
\eq
This means that the total population size $N$ will be dominated by $n_0$. Inserting $N=n_0$ into Eq. (\ref{eq:1}) we obtain that
\bq
	n_0 = N \approx f_0 K (1-\gamma-\mu),
\eq
and hence, for $i>0$,
\bq
	n_i \approx n_{i-1} \frac{f_{i-1}\mu/2}{(f_0 -f_i) (1-\gamma-\mu)} ,
\eq
and finally
\bq
	n_i \approx \frac{K\mu}{2\delta}\left(\frac{(1-\delta)\mu}{2\delta (1-\gamma-\mu)} \right)^{i-1}.
\eq
The above equation shows that (i) $n_i$ decreases exponentially with $i$, (ii) the rate of decrease is smaller for non-zero $\gamma$. This means that the abundance of genotype $L-1$ preceding the best-adapted genotype increases with increasing $\gamma$, even though the total population size decreases, as long as $L$ is sufficiently large, $\gamma$ is not too large, and our approximations remain valid. If the adaptation time $T$ is now limited by the last mutational step of going from $i=L-1$ to $i=L$, we can write that
\bq
	T = \frac{1}{\frac{\mu}{2}(1-\delta) n_{L-1}} = \frac{4\delta}{K\mu^2(1-\delta)}\left(\frac{2\delta (1-\gamma-\mu)}{\mu(1-\delta)} \right)^{L-2}. \label{eq:1dmodel}
\eq
Indeed, it turns out that the above formula agrees very well with computer simulations of the 1d model (Fig. \ref{fig:3}B), which supports our hypothesis that long fitness valleys slow down evolution in the full model, and that the presence of lethal genotypes facilitates valley crossing by increasing the abundance of less-fit but viable genotypes. 

It is interesting to compare this result with the observation of Ref. \cite{kuosmanen_turnover_2022} that decreased turnover increases the fixation probability of a mutation with a fixed selective advantage, thus decreasing the time to fixation if mutants occurs spontaneously and with a small rate. In our model, turnover, defined as the sum of per-capita birth and death rates, is indeed reduced for $\gamma>0$: for strain $i$, per-capita turnover is $f_i+N/K$, with $N$ decreasing with increasing $\gamma$. However, there is no fixation of valley genotypes in our model, so the result of Ref. \cite{kuosmanen_turnover_2022} does not directly apply. %Nevertheless, a classical result of population genetics shows that the equilibrium frequency of a deleterious mutation is $\mu/s$ where $s$ is the relative fitness disadvantage {\it per generation}. In our case, $s=(f_0-f_1)/f_0$   \red{where should turnover go here?}

%The latter must be caused by decreased selection, since the mutation rate $\mu$ is fixed while we vary $\gamma$.

% this section must be verified:
%Decreased selection in the presence of the lethal genotype reminisces  %Instead,  non-zero $\gamma$ decreases the ratio of net growth rates for adapted- and maladapted genotypes.  

\begin{figure}[t!]
	\begin{center}
	  \includegraphics[width=\columnwidth]{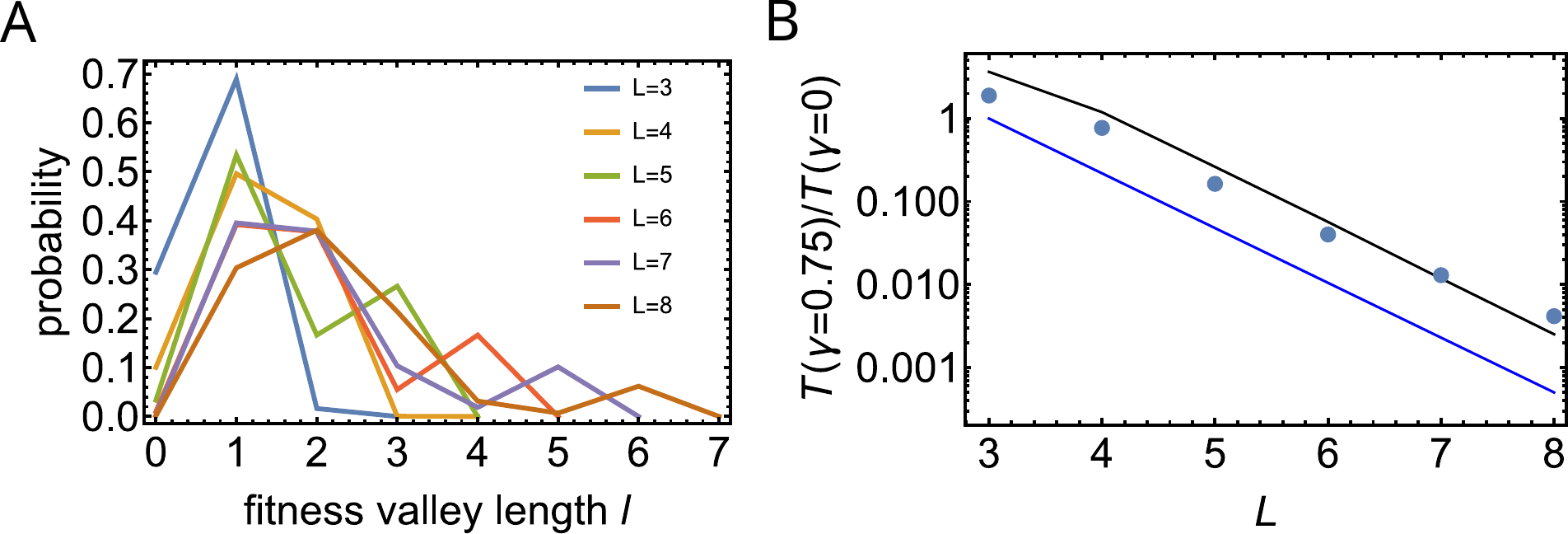}
		\caption{\label{fig:4}The linear-FL model predicts the adaptation time in the full model. (A) Probability of finding a fitness valley of length $l$, for different $L$. See the main text for the definition of $l$. (B) The ratio of adaptation times for the model with $\gamma=0.75$ and $\gamma=0$, for different $L$. Points = results from Fig. \ref{fig:1}, black line = Eq. (\ref{eq:Tav}), blue line = Eq. (\ref{eq:ratioT}). }
\end{center}
\end{figure}

We can now use the analytic expression (\ref{eq:1dmodel}) to make predictions for the full model with $2^L$ genotypes. First, we notice (Fig. \ref{fig:2}D) that for a fixed fitness landscape with a sufficiently deep and wide valley, evolution quickly reaches a local fitness maximum one mutation away from the initial genotype. To use Eq. (\ref{eq:1dmodel}) we must find the distribution $p_l$ of fitness valley length $l$. We can then calculate the ratio of the adaptation times for the model with and without lethal genotypes as follows:
\bq
    \sum_{l=0}^{L-1} T_l(\gamma) p_l \;/ \sum_{l=0}^{L-1} T_l(0) p_l .
    \label{eq:Tav}
\eq
To find $\{p_l\}$ numerically, we generated 10,000 random FLs, and found all fitness valleys, defined as consecutive runs of genotypes with monotonously decreasing fitness. Specifically, for each FL we found all such valleys along trajectories starting at the single-mutated genotype with maximum fitness and ending at the best adapted genotype. For each trajectory, we found the length (the number of links along which fitness decreases) of the shortest valley.   %  (For example, if fitness along a particular trajectory looks like this {0.99, 0.8, 0.6, 1} then the length of the valley will be 2.)   
We then took the minimum of all maximum lengths for each trajectory, and repeated this process for all trajectories and FLs. This gave the distribution of the minimum-length valley that evolution might encounter in our model. Such valleys would form a bottleneck limiting the rate of evolution on each specific FL.

Figure \ref{fig:4}A shows the distribution of fitness valley lengths for different sizes $L$ of the FL. Using the numerically obtained $p_l(L)$, we calculated the ratio (\ref{eq:Tav}) for different $L$, assuming $\delta=0.3$ (average depth of fitness valley from Fig. (\ref{fig:2})). Figure \ref{fig:4}B shows that this estimate approximately agrees with the ratio obtained from the adaptation times from Fig. \ref{fig:1}.

Interestingly, an even simpler approach amounting to replacing $L \to L-1$ in Eq. (\ref{eq:1dmodel}) and calculating the ratio of adaptation times as
%Treating this local maximum as genotype $i=0$ in the 1d model, we replace $L\to L-1$ in Eq. (\ref{eq:1dmodel}) and obtain the following ratio of adaptation times for the models with $\gamma>0$ and  $\gamma=0$, both following the same trajectory:
\bq
	\frac{T_{L-1}(\gamma>0)}{T_{L-1}(\gamma=0)} = \left(\frac{(1-\gamma-\mu)}{(1-\mu)}
 \right)^{L-3} \label{eq:ratioT}
\eq
qualitatively reproduces the data (Fig. \ref{fig:4}B). This is because, as seen in Fig. \ref{fig:4}A, a fitness landscape for binary sequences of length $L$ always has a non-zero probability of having a fitness valley of length $L-2$. These longest valleys dominate the adaptation time due to its exponential dependence on the fitness valley length. Even though the probability $p_{L-2}$ decreases exponentially with $L$ as $\sim \exp(-0.5 L)$ (Fig. \ref{fig:4}A), the rate of decrease is not sufficient to overcome the exponential increase of $T_l(\gamma)$  with rate $2\delta (1-\gamma-\mu)/(\mu(1-\delta))$ as long as $\mu$ is sufficiently small.
%where $l$ is the length of the easiest-to-traverse fitness valley ($l$ equals the number of genotypes, peak-to-peak). 
However, due to the presence of shorter fitness valleys in many landscapes (Fig. \ref{fig:4}A), the lethal-genotype model will perform better only on a subset of FLs which do not have such short valleys.

%Equation (\ref{eq:ratioT}) shows that %making $\gamma$ as close to $1-\mu$ as possible will lead to the largest differences between the models. On the other hand, too large $\gamma$ will cause the population size to become small and the population may be prone to a stochastic exctinction \cite{ref}.
%Second, 
%the ratio of adaptation times increases exponentially with the length $l$ of the valley. If $l$ increases with $L$, then the lethal-genotype model will always be faster for sufficiently large random fitness landscape, not only on a subset of such landscapes as we have showed.

%Let us estimate the expected length of the shortest valley for a random fitness landscape of size $2^L$.
%CAN IT BE DONE?

%-------------Discussion----------------
To summarize, we have observed a significant reduction of the adaptation time in the presence of lethal mutations. The effect is not caused by a reduced number of evolutionary accessible paths; unlike in the case of ``holey landscapes'' \cite{gavrilets_percolation_1997,gavrilets_dynamical_1999}, lethal genotypes in our model are external to the hypercube fitness landscape and increasing $\gamma$ does not affect the structure of the FL.
The effect is also not due to evolution avoiding fitter genotypes for the sake of less-adapted ones, or vice-versa; we have seen that evolutionary trajectories are similar for both models.
Rather, the presence of lethal genotypes decreases the effective depth of fitness valleys by reducing selection against less-fit genotypes. This increases the abundance of valley genotypes and hence the rate with which best-adapted mutants are generated. 

It would be interesting to see if our conclusions hold for fitness landscapes more realistic than the maximally-random (``House-of-Cards'') FL used here \cite{de_visser_empirical_2014}, fitness landscapes that evolve over time, and non logistic-growth like growth models, in particular for slowly expanding populations such as bacterial colonies \cite{hallatschek_life_2010}. Moreover, in the sequential fixation regime (low mutation regime, small fitness differences) evolution can be faster for smaller population sizes \cite{jain_evolutionary_2011}, which would be of interest to check as well.
Our model could be also extended to the case of ``dormant states'' which do not replicate but can be transformed back into viable states in a way similar to what happens in ``gene bank'' models \cite{lennon_principles_2021}.

{\it Data and code}. The software required to reproduce the results presented here can be found at \url{https://github.com/Dioscuri-Centre/quasilethal}.

 {\it Acknowledgments. } We acknowledge funding under Dioscuri, a programme initiated by the Max Planck Society, jointly managed with the National Science Centre in Poland, and mutually funded by Polish Ministry of Science and Higher Education and German Federal Ministry of Education and Research (UMO-2019/02/H/NZ6/00003). V.B. is grateful for support from the U.S. National Academy of Sciences (NAS) and the Polish Academy of Sciences (PAS) for scientists from Ukraine.

%of course in reality the fitness landscape won't be 0..1
% and fitness may continue to increase more gradually than jumping to approx 1 after the first step

%\red{BW finished here}

%Our prediction could be tested in a laboratory experiment using a population of bacteria in a continuous-culture bioreactor.  HOW? (Large landscape required with similar fitness and the beginning and the end).

%\ba
%	n_i \to n_i+1 \; \mbox{with rate} n_i(1-\mu-\gamma), \\
%	n_i \to n_i+1 \; \mbox{with rate} n_i(1-\mu-\gamma), \\
%\ea

\bibliographystyle{apsrev-nourl}
\bibliography{literature}

\end{document}